\documentclass[a4paper,11pt]{article}
\usepackage{pos}
\usepackage{graphicx}          %Include figure files
\usepackage{hyperref}

\title{Recent Results from the FASTSUM Collaboration}
\author*[a]{Chris Allton}
\author[a,b]{Gert Aarts}
\author[a]{Ryan Bignell}
\author[a]{Tim Burns}
\author[a]{Sergio Chaves}
\author[c]{Simon Hands}
\author[d]{Benjamin J\"ager}
\author[e]{Seyong Kim}
\author[f]{Maria Paola Lombardo}
\author[a]{Ben Page}
\author[g]{Sin\'ead M. Ryan}
\author[h]{Jon-Ivar Skullerud}
\author[a]{Thomas Spriggs}

\onbehalf{for the FASTSUM collaboration}

\affiliation[a]{Department of Physics, Swansea University, Swansea,
  SA2 8PP, United Kingdom}

\affiliation[b]{European Centre for Theoretical Studies in Nuclear Physics and Related
Areas (ECT*) Fondazione Bruno Kessler, Strada delle Tabarelle 286,
38123 Villazzano (TN), Italy}

\affiliation[c]{Department of Mathematical Sciences, University of
Liverpool, Liverpool L69 3BX, United Kingdom}

\affiliation[d]{CP3-Origins \& Danish IAS, Department of Mathematics and Computer
Science, University of Southern Denmark, 5230, Odense M, Denmark}

\affiliation[e]{Department of Physics, Sejong University, Seoul 143-747, Korea}

\affiliation[f]{INFN, Sezione di Firenze, 50019 Sesto Fiorentino (FI), Italy}

\affiliation[g]{School of Mathematics, Trinity College, Dublin, Ireland}

\affiliation[h]{Department of Theoretical Physics, National University
  of Ireland Maynooth, County Kildare, Ireland}

\emailAdd{c.r.allton@swansea.ac.uk}

\abstract{The {\sc Fastsum} Collaboration has developed a
  comprehensive research programme in thermal QCD using $2+1$ flavour,
  anisotropic ensembles. In this talk, we summarise some of our recent
  results including thermal hadron spectrum calculations using our
  ``Generation 2L'' ensembles which have pion masses of 239(1)
  MeV. These include open charm mesons and charm baryons. We also
  summarise our work using the Backus Gilbert approach to determining
  the spectral function of the NRQCD bottomonium system.  Finally, we
  review our determination of the interquark potential in the same
  system, but using our ``Generation 2'' ensembles which have heavier
  pion masses of 384(4) MeV.  }

\FullConference{%
The 39th International Symposium on Lattice Field Theory (Lattice
2022), \\
8-13 August, 2022\\
Bonn, Germany
}

\begin{document}
\maketitle

%}}}

%{{{ Introduction

\section{Introduction}

The thermal spectrum of QCD is of great interest for intrinsic reasons
in order to understand how confinement is manifest in hadronic
systems.  It is also crucial to aid the analysis of heavy-ion
collision experiments.  Here we present an update of the {\sc Fastsum}
Collaboration's thermal hadronic spectrum research, focussing on
open charm mesons, charm baryons and bottomonium. We
also present an update of our studies of the interquark potential in
bottomonium.

We use 2+1 flavour dynamical simulations with anisotropic latices
where the temporal lattice spacing, $a_\tau$, is smaller than the spatial
one, $a_s$ \cite{Aarts:2014nba,Aarts:2020vyb}. Our anisotropy is
designed to maximise information from thermal temporal correlators,
noting that they are constrained in temporal extent, $L_\tau$, since
the temperature, $T = 1/L_\tau$. We use stout-linked, clover-improved
Wilson fermions and Symanzik-improved gauge fields.

The lattice ensembles used in this work are our Generation 2 and 2L
ensembles which have parameters listed in Tables \ref{tab:gen2}
and \ref{tab:gen2l}. These ensembles have a pion mass of 384(4) and
239(1) MeV respectively and span temperatures both below and above the
pseudocritical temperature $T_\text{pc}$.
\bigskip

\begin{table}[h]
  \begin{center}
\begin{tabular}{l|rrrr|rrrr}
$N_\tau$             & 128 & 40  & 36  & 32  & 28  & 24  &  20 & 16 \\ \hline
$T\,\, \text{[MeV]}$ &  44 &141  &156  &176  &201  &235  &281  &352
\end{tabular}
  \end{center}
\caption{Parameters for the {\sc Fastsum} Generation 2 ensembles used
  in this work. The lattice sizes are $24^3 \times N_\tau$, with
  lattice spacings $a_s = 0.1205(8)$ fm and $a_\tau = 35.1(2)\,$am,
  and pion mass $m_\pi = 384(4)$ MeV. The vertical line indicates the
  position of $T_\text{pc}\approx 181$ MeV. Full details in
  \cite{Aarts:2014nba,Aarts:2020vyb}.}
\label{tab:gen2}
\end{table}

\begin{table}[h]
  \begin{center}
    \begin{tabular}{l|rrrrr|rrrrrr}
$N_\tau$ & 128 & 64 & 56 & 48 & 40 & 36 & 32 & 28 & 24 & 20 & 16 \\ \hline
$T\,\, \text{[MeV]}$ & 47 & 95 & 109 & 127 & 152 & 169 & 190 & 217 & 253 & 304 & 380
\end{tabular}
  \end{center}
\caption{Parameters for the {\sc Fastsum} Generation 2L ensembles used
  in this work. The lattice sizes are $32^3 \times N_\tau$, with
  lattice spacings $a_s = 0.1121(3)$ fm and $a_\tau = 32.46(7)\,$am,
  and pion mass $m_\pi = 239(1)$ MeV~\cite{Wilson:2019wfr}. The
  vertical line indicates the position of $T_\text{pc}\approx 167$
  MeV. Full details in \cite{Aarts:2020vyb,Aarts:2022krz}.}
\label{tab:gen2l}
\end{table}

%}}}

%{{{ Charm hadron spectrum

\section{Charm hadron spectrum}

Here we summarise results from both open-charm mesons and charmed
baryons using our Generation 2L ensembles.

Unlike hiddened-charmed mesons at non-zero temperature, which have been
extensively studied on the lattice \cite{Rothkopf:2019ipj},
open-charmed mesons have received less attention
\cite{Bazavov:2014yba, Bazavov:2014cta,Kelly:2018hsi}. In
\cite{Aarts:2022krz}, we present results for the open charm meson
spectrum for $T \lesssim T_{pc}$.  Because the states are confined, we
proceed with conventional analysis techniques and assess up to which
temperatures these are applicable. We extend these techniques to
determine the variation of the masses as a function of temperature.

\begin{figure}[h]
\begin{center}
  \includegraphics[width=0.48\textwidth]{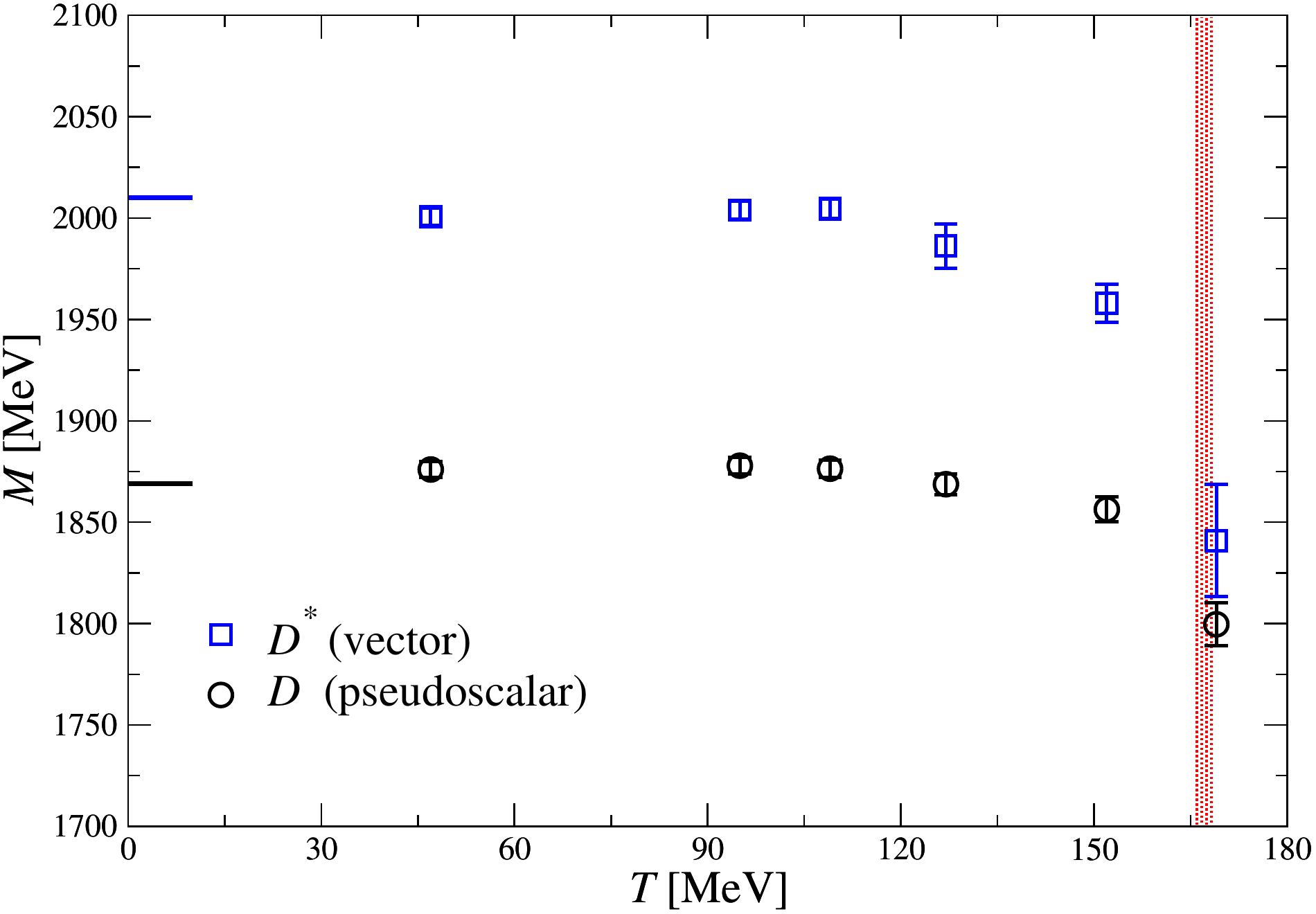}
  \includegraphics[width=0.48\textwidth]{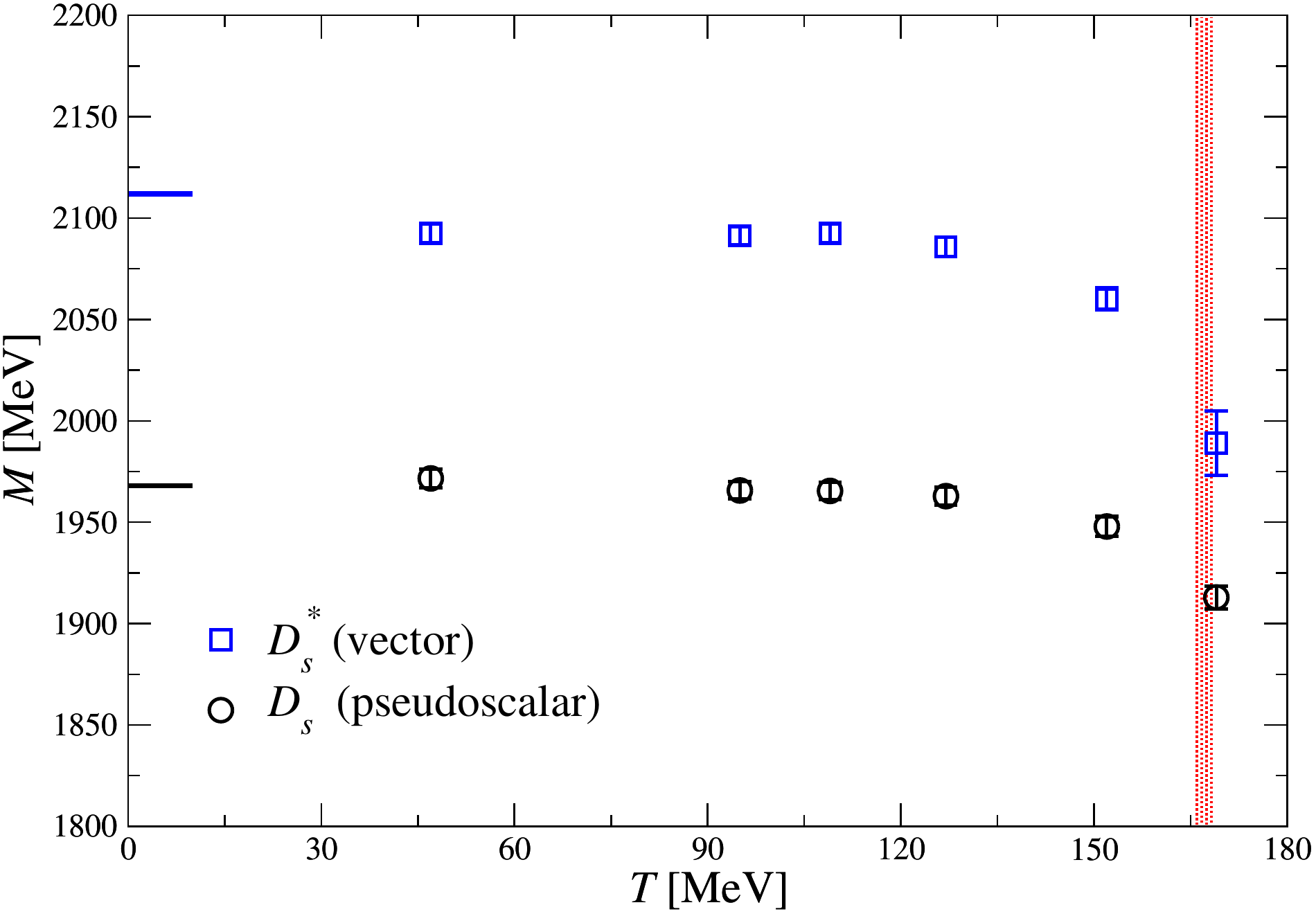}
\end{center}
\caption{Temperature dependence of the groundstate masses in the
  hadronic phase, for $D$ and $D^*$ (left) and $D_s$ and $D_s^*$
  (right) mesons. The vertical band indicates $T_{pc}$ and
  the horizontal stubs at $T = 0$ represent the PDG values
  \cite{ParticleDataGroup:2020ssz}.}
\label{fig:charm-mesons}
\end{figure}

The results are shown in Fig.\ref{fig:charm-mesons}.  These show a
small temperature variation where both the pseudoscalar,
$D_{\rm{(s)}}$, and vector, $D^\ast_{\rm{(s)}}$, mesons' masses
decrease as the temperature approches $T_{pc}$. However, note that
this temperature shift is at the percent level.
In contrast, the analogous thermal effects for the axial vector and scalar
channel are very strong, see \cite{Aarts:2022krz} for details.

In another analysis \cite{Bignell:lat22, charm-baryons}, we study the
charm baryon spectrum paying particular attention to both parity
states.  We extract the masses in the confined phase and use a method
based on a direct analysis of the correlation function to determine
whether the parity states approach degeneracy for $T\ge T_{pc}$.

\begin{figure}[h]
\begin{center}
  \includegraphics[width=0.49\textwidth]{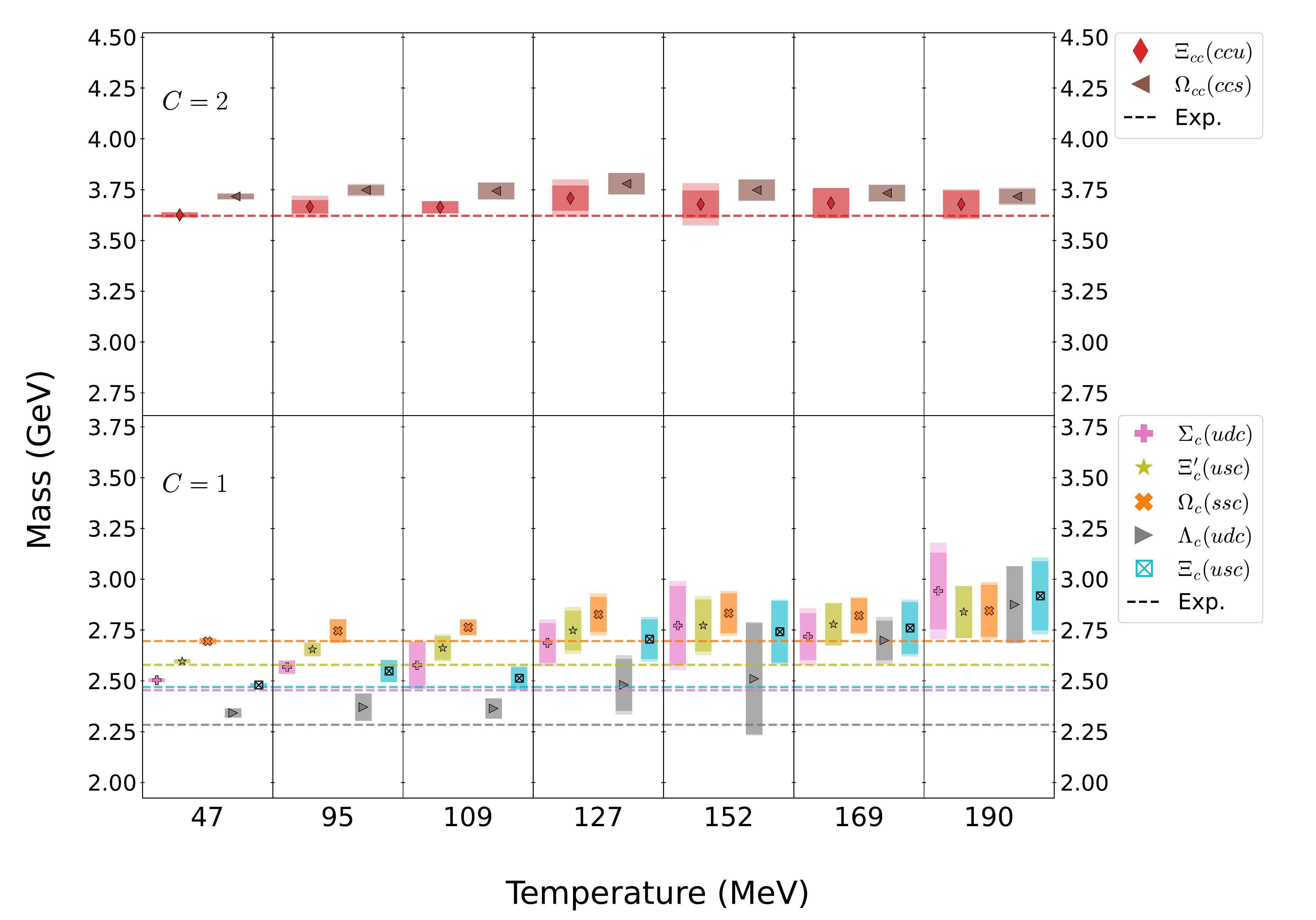}
  \includegraphics[width=0.49\textwidth]{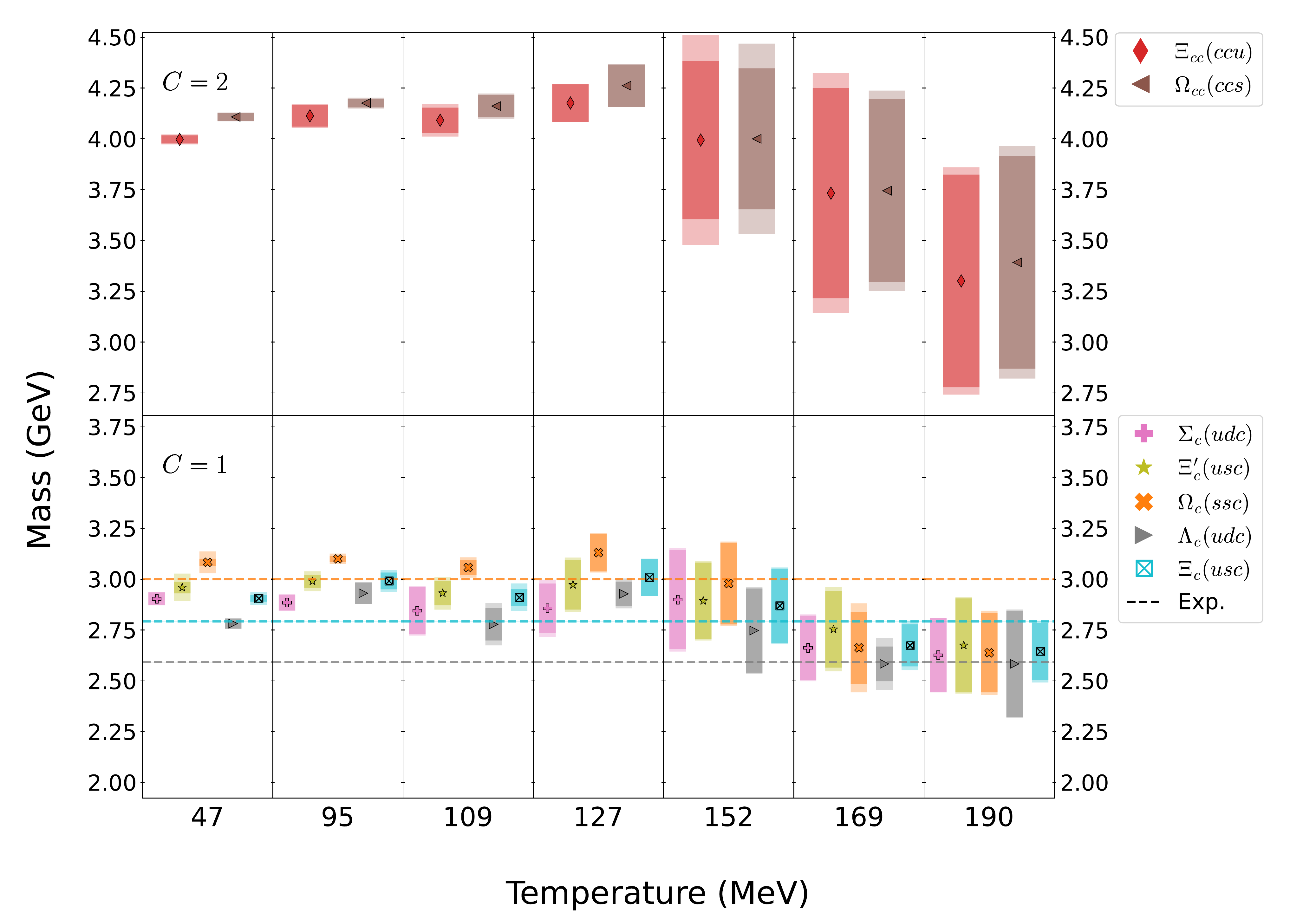}
\end{center}
  \caption{Mass spectrum of of the $J^{1/2}{}^{+}$ baryons with
    positive (left) and negative parity (right) as a function of
    temperature. Dashed lines are zero-temperature experimental
    results~\cite{ParticleDataGroup:2020ssz} to guide the eye. The
    inner (outer) shaded regions represents the statistical
    (systematic) errors. See \cite{Bignell:lat22} for
    the corresponding plots for the $J^{3/2}{}^{+}$ states.}
\label{fig:charm-baryons-mass}
\end{figure}

In Fig.\ref{fig:charm-baryons-mass} we plot the masses for baryons
with a variety of charm content as a function of $T$ up to just beyond
$T_{pc}$ where the simple pole fits become unreliable.

\begin{figure}[h]
\begin{center}
  \includegraphics[width=0.60\textwidth]{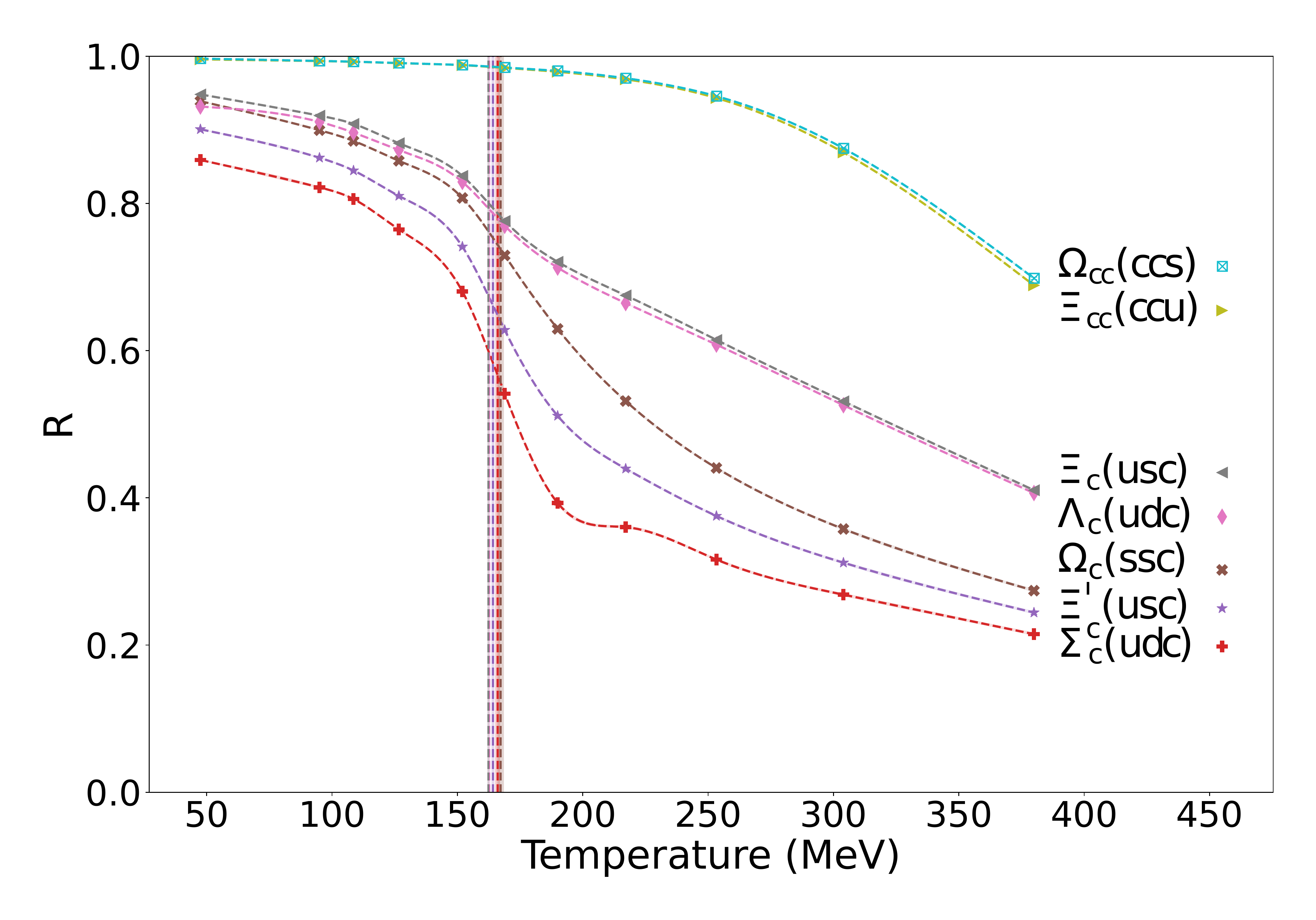}
\end{center}
\vspace*{-5mm}
\caption{ R value from Eq.(\ref{eq:R}) for $J=\frac{1}{2}$ baryons
  with the lines from cubic splines. The transition temperature
  estimates obtained from the inflection point of the splines are
  shown as vertical lines.}
\label{fig:charm-baryons-R}
\end{figure}

For $T>T_{pc}$ we cannot assume that the charm baryon states are bound
and so a conventional pole fitting ansatze cannot be applied. To gain
information about the mass of the two parity
states, we therefore define the ratio $R$,
\begin{equation}
R(n_0) = \frac{
  \sum_{n=n_0}^{\frac{1}{2}N_\tau -1}\,
      \mathcal{R}(\tau_{n})/\sigma_{\mathcal{R}}^2(\tau_{n})}
 {\sum_{n=n_0}^{\frac{1}{2}N_\tau -1}\,
      1                    /\sigma_{\mathcal{R}}^2(\tau_{n})},
\;\;\;\;\;\;\;\;\;
    {\rm where}\;\;\;\;\;
\mathcal{R}(\tau) = \frac{G^{+}(\tau) -G^{-}(\tau)}{G^{+}(\tau) + G^{-}(\tau)}.
\label{eq:R}
\end{equation}
%
%\begin{align}
%\mathcal{R}(\tau) &= \frac{G^{+}(\tau) -G^{-}(\tau)}{G^{+}(\tau) + G^{-}(\tau)} \\
%R(n_0) &= \frac{
%  \sum_{n=n_0}^{\frac{1}{2}N_\tau -1}\,
%      \mathcal{R}(\tau_{n})/\sigma_{\mathcal{R}}^2(\tau_{n})}
% {\sum_{n=n_0}^{\frac{1}{2}N_\tau -1}\,
%      1                    /\sigma_{\mathcal{R}}^2(\tau_{n})}.
%\end{align}
%
Typically we use $n_0=4$, and our results are not qualitatively
sensitive to this choice.  $R$ will be unity in the limit of $M^+ \ll
M^-$ and zero for the degenerate case. In
Fig.\ref{fig:charm-baryons-R} we plot $R$ for a number of channels. We
can see an approach to degeneracy above $T_{pc}$ which is most
pronounced for baryons with the least charm content.  By fitting the
data to cubic splines, we can determine estimates of the transition
temperatures from the inflection points and these are indicated as
vertical lines in the figure.  We note that the location of the
inflection points coincide, within a few MeV, with the pseudocritical
temperature, $T_\text{pc}=167(3)$ MeV, determined via the chiral
condensate \cite{Aarts:2022krz} and hence are a manifestion of chiral
symmetry restoration in the charmed baryon sector. Further details
about these points are elucidated in \cite{Bignell:lat22,charm-baryons}.

%}}}

%{{{ Bottomonium (NRQCD) spectrum

\section{Bottomonium (NRQCD) spectrum}

Bottomonium states are important probes of the quark gluon plasma
phase in heavy-ion collision experiments because they are created very
early and do not reach chemical equilibrium.  The lattice approach to
analysing bottomonium spectra invariably relies on extracting the
spectral function at temperature $T$, $\rho(\omega,T)$, defined from
the correlation function,
\begin{equation}
G(\tau;T) = \int_{\omega_\text{min}}^\infty \frac{d\omega}{2\pi} K(\tau,\omega) \rho(\omega;T),
\label{eq:spectral}
\end{equation}
where the kernel for NRQCD quarks is defined as,
\begin{equation}
K(\tau,\omega) = e^{-\omega \tau}.
\end{equation}
Note that since NRQCD introduces an additive energy shift, the lower
limit of the integral in Eq.(\ref{eq:spectral}), $\omega_\text{min}$,
is not necessarily zero.
The spectral function gives complete information about the spectrum of
a particular channel, including the widths of the states.
The {\sc Fastsum} Collaboration has studied the bottomonium spectrum
using NRQCD bottom quarks in a number of publications, e.g.
\cite{Aarts:2014cda}, using a variety of methods to extract
the spectral function.

In \cite{Page:lat22} we extend this work by using the Backus Gilbert
\cite{backus gilbert} method to obtain the spectral function with our
Generation 2L ensembles, and we report on this work here.  We note
that we can introduce two ``hyper-parameters'' in our analysis. The
first is the ``whitening'' factor, $\alpha$, in the Tikhonov-like
method \cite{tikhonov}, which governs how much of the identity (white
noise) is added to the kernel. To remove the $\alpha$ dependency in
the final result, the $\alpha \rightarrow 0$ limit is taken. The
second parameter is an energy shift, $\Delta$. Since
Eq.(\ref{eq:spectral}) is a Laplace transform, we can trivially shift
$\rho(\omega) \rightarrow \rho(\omega+\Delta)$ by multiplying
$G(\tau)$ by $e^{\Delta \tau}$. Increasing the energy shift, $\Delta$,
moves the ground state feature closer to $\omega_\text{min}$ which is
advantageous because that is where the Backus Gilbert method has the
greatest resolution. Note however that the value of $\Delta$ needs to
be limited to ensure that no spectral feature is pushed into the
$\omega<\omega_\text{min}$ region.  Hence we remove the dependency on
the $\Delta$ hyper-parameter via this requirement.

\begin{figure}[h]
\begin{center}
  \includegraphics[width=0.70\textwidth]{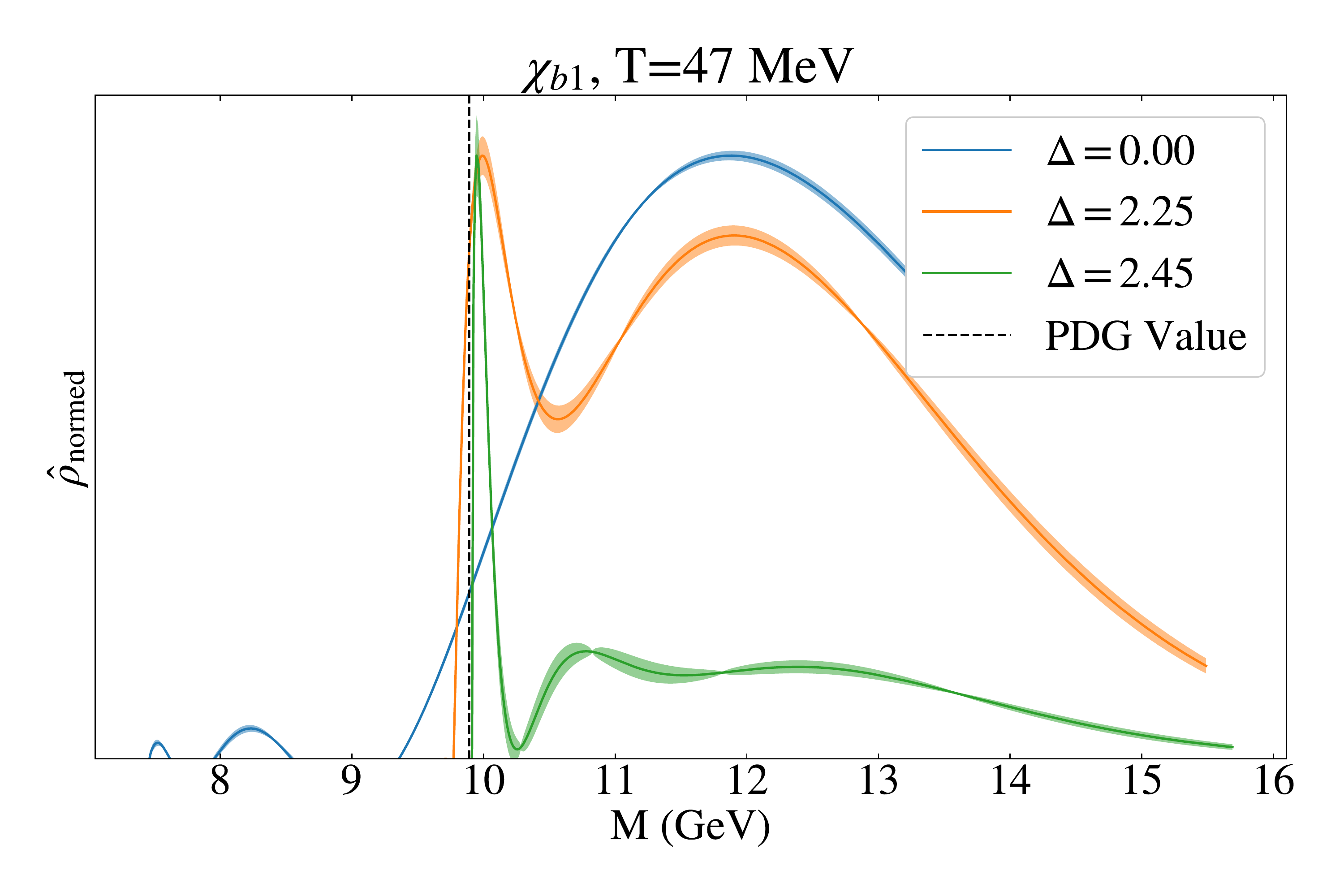}
\end{center}
\vspace*{-5mm}
\caption{The $\chi_{b1}$ spectral function obtained via the Backus
  Gilbert method for a variety of $\Delta$ energy shifts at $T=47$
  MeV.}
\label{fig:bottomonium-BG}
\end{figure}

In Fig.\ref{fig:bottomonium-BG} we plot the $\chi_{b1}$ spectral
function obtained from local correlators with various $\Delta$ values
to illustrate that the ground state feature becomes better resolved as
$\Delta$ increases. Full results and predictions of masses and widths
obtained using this method are in \cite{Page:lat22}.

We can also perform an interesting statistical analysis of the
correlation functions. As pointed out by Parisi and Lepage \cite{Parisi:1983ae,
  Lepage:1989hd}, the statistical error of the hadronic correlation
function, $\langle {\cal O}(t) {\cal O}(O) \rangle$ at large time is
determined by the lightest states that can be composed from ${\cal
  O}^2$. Typically this will be the pseudoscalar state.

We have analysed the statistical error in the bottomonium correlation
functions by measuring their covariance matrices' singularity as the
energy shift, $e^{\Delta \tau}$, is applied. The value of $\Delta$
corresponding to the most singular covariance matrix,
$\Delta_{\rm{sing}}$, is a prediction of (half) the ground state mass in the
${\cal O}^2$ channel.
We used the condition number of the covariance
matrix to determine $\Delta_{\rm{sing}}$. Following the Parisi and
Lepage analysis, we expect to find $\Delta_{\rm{sing}} = M_{\eta_b}$
i.e. the mass of the pseudoscalar in the bottomonium sector.

\begin{figure}[h]
  \centering
  \includegraphics[width=0.7\textwidth]{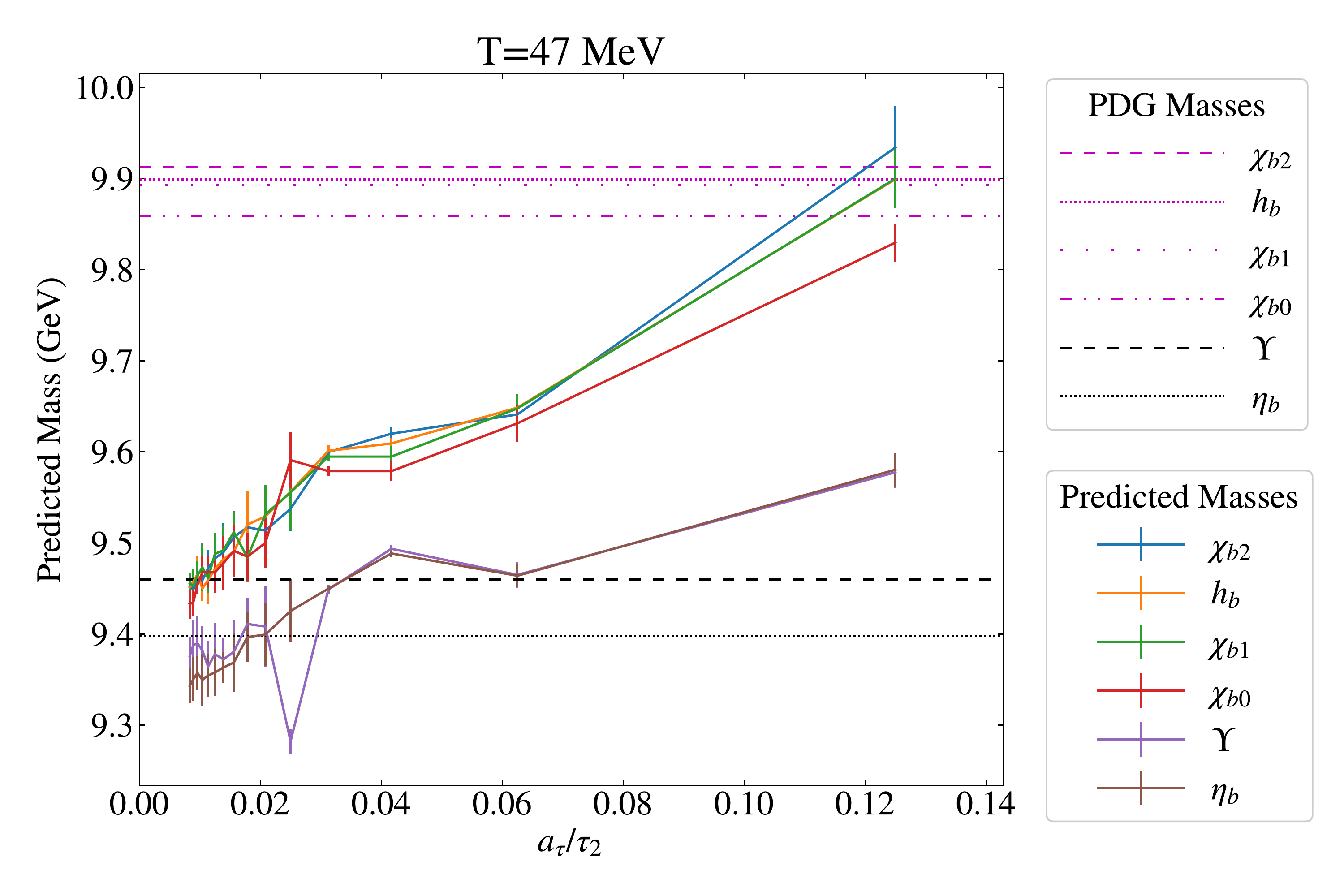}
\caption{The value of the energy shift, $\Delta_\text{sing}$, (i.e. the predicted
  mass) which gives the most singular covariance matrix for a variety
  of bottomonium channels as a function of $1/\tau_2$. The covariance
  matrices are defined over the time interval $[0,\tau_2/a_\tau]$ and
  therefore the best results are obtained as $\tau_2 \rightarrow
  \infty$. Experimental mass values are also shown as horizontal lines
  \cite{ParticleDataGroup:2020ssz}. This indicates the method
  correctly predicts the $\eta_b$ (i.e. the pseudoscalar) mass as
  $\tau_2$ increases.}
\label{fig:bottomonium-parisi-lepage}
\end{figure}

In Fig.\ref{fig:bottomonium-parisi-lepage} we plot results for the
channels considered ($\eta_b, \Upsilon, h_b$ and
$\chi_{\rm{b0},\rm{b1},\rm{b2}}$) as a function of $\tau_2$ where the
covariance matrices are analysed over the time interval
$[0,\tau_2/a_\tau]$. Smeared operators at both the source and sink
were used. Figure \ref{fig:bottomonium-parisi-lepage} shows a
convergence, as $\tau_2$ increases, to a mass value compatible with
the pseudoscalar, $\eta_b$, independent of the channel, as expected
from an analysis following \cite{Parisi:1983ae, Lepage:1989hd}.
Further details of this work are discussed in \cite{Page:lat22}.

%}}}

%{{{ Interquark potentials in Bottomonium

\section{Interquark potentials in Bottomonium}

For temperatures below $T_{pc}$, the interquark potential is known to
be confining, i.e. with a non-zero string tension, whereas above
$T_{pc}$ it is expected to flatten allowing unbound quark states.
Lattice studies of the interquark potential have been obtained from
Wilson loops, which correspond to infinitely heavy quarks
\cite{Burnier:2014ssa, HotQCD:2014kol}, the NRQCD bottomonium system
\cite{Larsen:2020rjk}, and from charmonium using the HAL QCD method
\cite{Kawanai:2011jt,Evans:2013yva,Allton:2015ora}.

In this work, we study the interquark potential in the bottomonium system
also using the HAL QCD approach, see \cite{Spriggs:lat22}.
The bottom quarks were simulated using the NRQCD approximation and our
Generation 2 ensembles were used, see Table \ref{tab:gen2}.

The HAL QCD method \cite{Aoki:2012tk} uses Bethe Salpeter
wavefunctions, $\psi(t,\vec{r})$, obtained from temporal correlators
of non-local heavy quark--antiquark meson operators, $\overline{Q}(\tau,\vec{x})
\Gamma Q(\tau,\vec{x}+\vec{r})$. It represents the mesonic system with a
Schr\"odinger equation,
\begin{equation}
\left[ \frac{p^2}{2\mu} + V(\vec{r}) \right]\psi(\tau,\vec{r}) = E \psi(\tau,\vec{r}),
\end{equation}
where $\mu$ is the reduced quark mass in the centre of mass frame.
The residual, non-physical $\tau-$dependency has to be carefully handled
and this is discussed in \cite{Spriggs:lat22}.

\begin{figure}[h]
\begin{center}
  \includegraphics[width=0.49\textwidth]{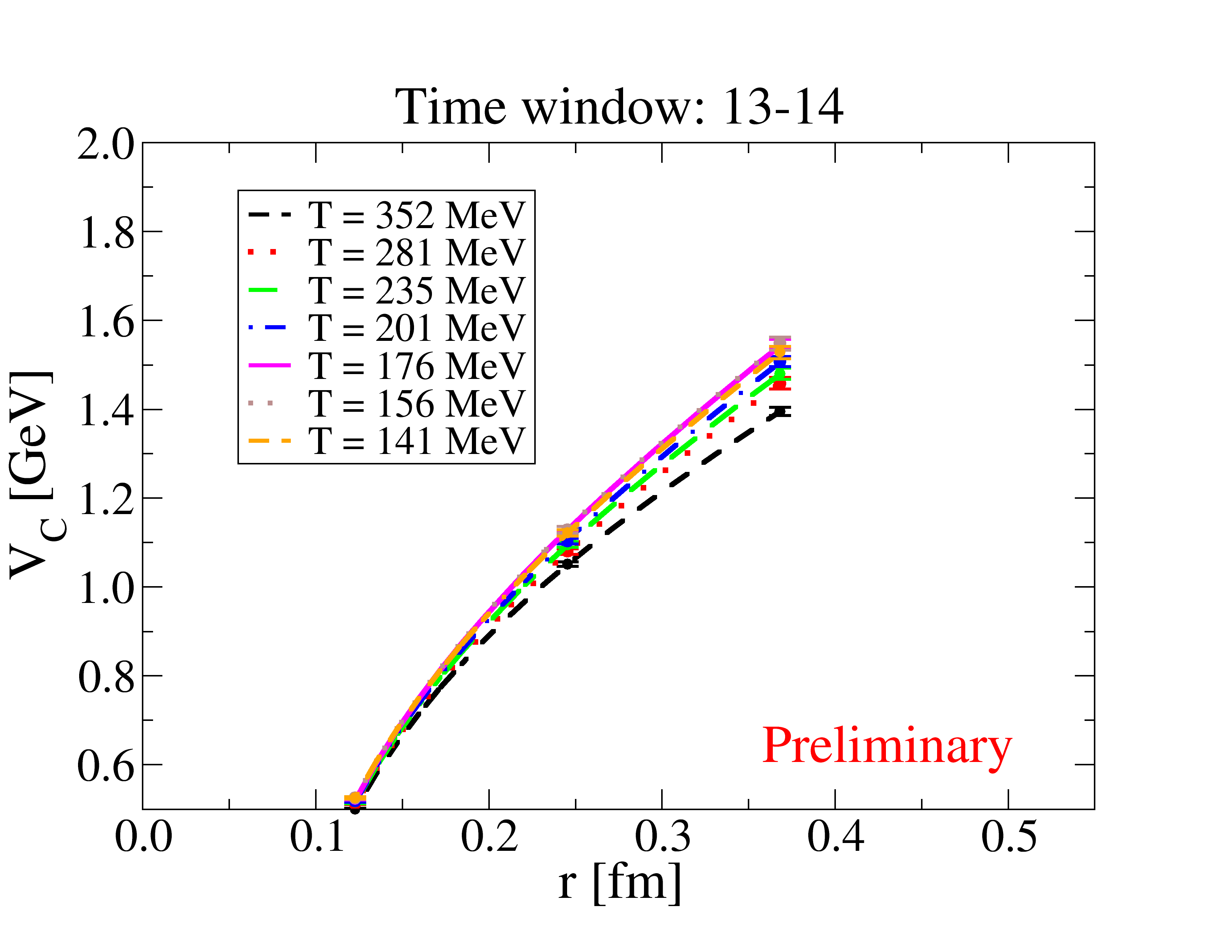}
  \includegraphics[width=0.49\textwidth]{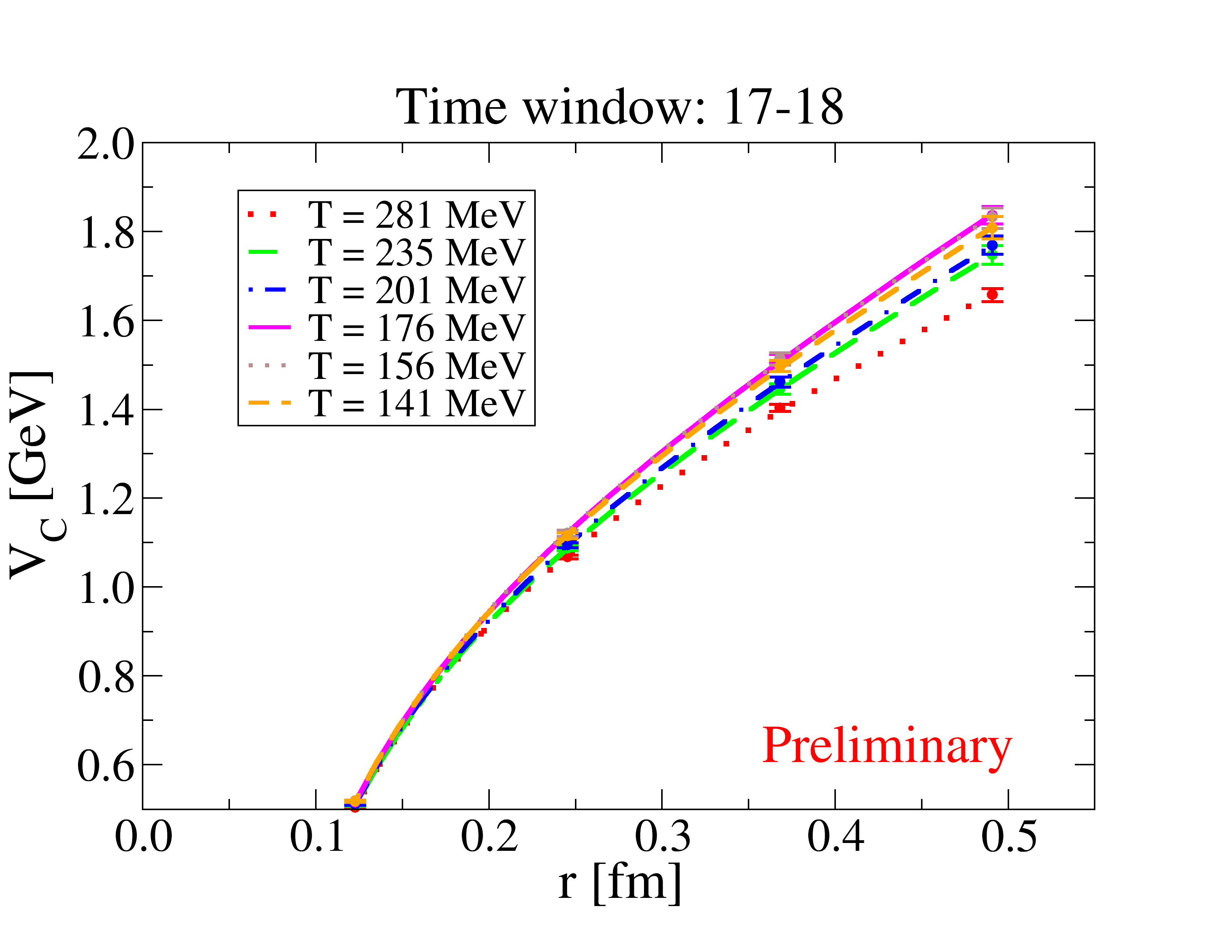}
\end{center}
\hspace*{-5mm}
\caption{The interquark potential in (NRQCD) bottomonium via the HAL
  QCD procedure. In each pane, all the temperatures used the same time
  window: $[13,14]$ (left) and $[17,18]$ (right). This is to isolate
  thermal effects from $\tau$ systematics. The expected flattening of
  the potential with temperature can be seen.}
\label{fig:potential}
\end{figure}

Figure \ref{fig:potential} shows the preliminary results for the
interquark bottomonium potential for a variety of temperatures.  In
each pane, the {\em same} time window was considered, thus nullifying
any systematic effects from this fitting artefact.  These plots show
evidence of the expected flattening of the potential as the
temperature increases.

%}}}

%{{{ Conclusions

\section{Conclusions}

Recent results from {\sc Fastsum} Collaboration's thermal spectrum
research \cite{Aarts:2022krz,Bignell:lat22,charm-baryons,Page:lat22}
and interquark potentials \cite{Spriggs:lat22} have been presented.

%}}}

%{{{ Acknowledgements

\section{Acknowledgements}

This work is supported by STFC grant ST/T000813/1.
SK is supported by the National Research Foundation of Korea under
grant NRF-2021R1A2C1092701 and Grant NRF-2021K1A3A1A16096820, funded
by the Korean government (MEST).
BP has been supported by a Swansea University Research Excellence
Scholarship (SURES).
This work used the DiRAC Extreme Scaling service at the University of
Edinburgh, operated by the Edinburgh Parallel Computing Centre and the
DiRAC Data Intensive service operated by the University of Leicester
IT Services on behalf of the STFC DiRAC HPC Facility
(www.dirac.ac.uk). This equipment was funded by BEIS capital funding
via STFC capital grants ST/R00238X/1, ST/K000373/1 and ST/R002363/1
and STFC DiRAC Operations grants ST/R001006/1 and ST/R001014/1. DiRAC
is part of the UK National e-Infrastructure.
This work was performed using PRACE resources at Cineca (Italy), CEA
(France) and Stuttgart (Germany) via grants 2015133079, 2018194714,
2019214714 and 2020214714.
We acknowledge the support of the Swansea Academy for Advanced
Computing, the Supercomputing Wales project, which is part-funded by
the European Regional Development Fund (ERDF) via Welsh Government,
and the University of Southern Denmark and ICHEC, Ireland for use of
computing facilities.
We are grateful to the Hadron Spectrum Collaboration for the use of
their zero temperature ensemble in our Generation 2 work.

%}}}

%{{{ Bibliography

%}}}

\end{document}